# Non-Hermitian approach for quantum plasmonics

Cristian L. Cortes, Matthew Otten and Stephen K. Gray

Center for Nanoscale Materials

Argonne National Laboratory

Lemont, Illinois 60439

**Abstract**

We examine the limits of applicability of a simple non-Hermitian model for exciton/plasmon interactions in the presence of dissipation and dephasing. The model can be used as an alternative to the more complete Lindblad density matrix approach and is computationally and conceptually simpler. We find that optical spectra in the linear regime can be adequately described by this approach. The model can fail, however, under continuous optical driving in some circumstances. In the case of two quantum dots or excitons interacting with a plasmon, the model can also describe coherences and entanglement qualitatively when both dissipation and dephasing are present, and quantitatively in the limit with no dephasing. The model can also be extended to the case of many quantum dots interacting with a plasmon and results, within the single-excitation manifold, are presented for the fifty quantum dot case.

## I. INTRODUCTION

Plasmonics is the study of light interactions with generally metallic structures that can be resonantly excited to yield intense and localized electromagnetic responses of interest both for fundamental and practical reasons;  see, e.g., Refs. [1] and [2].  Of interest is the coupling of plasmonic systems with molecules, nanostructures or materials that exhibit quantum mechanical responses that could conceivably be enhanced or coupled into the plasmonic structure in some fashion to achieve interesting outcomes, i.e. quantum plasmonics [3,4].

We have previously modeled quantum dot/plasmon interactions [5-7] using a cavity quantum electrodynamics approach based on the Lindblad master equation [8] for the quantum mechanical density matrix.  This is a reasonably rigorous approach  which can incorporate important environmental effects such as dissipation and dephasing  but can be computationally intensive.  Furthermore, it is desirable to develop simpler models for analysis purposes that can still convey some of the correct dynamics.

Non-Hermitian quantum mechanics generally corresponds to the study of time-dependent or time-independent Schrödinger equations with Hamiltonian operators that are not Hermitian [9, 10].  The  non-Hermitian  terms in the Hamiltonian are designed to describe processes such as interaction with an environment that are not explicitly included as degrees of freedom  in the Hamiltonian.  While much recent work has focused on Parity-Time (PT) symmetric non-Hermitian Hamiltonians that have real eigenvalues and can exhibit interesting properties such as exceptional points [10], our focus here is on  non-Hermitian Hamiltonians with complex eigenvalues that can mimic to some degree dissipation and dephasing relevant to quantum dot or exciton interactions with plasmonic systems such as metal nanoparticles or arrays of metal nanoparticles.  The non-Hermitian Hamiltonian results in systems of equations that qualitatively (and sometimes quantitatively) capture the true dynamics while being significantly more efficient to solve compared to the full density matrix approach. Using our non-Hermitian Hamiltonian can allow the study of systems far larger than otherwise possible, such as the investigation of entanglement dynamics of large numbers of quantum dots coupled to a plasmonic system.

Section II below outlines both the Lindblad (IIA) and non-Hermitian (IIB) approaches, Sec. III outlines the results we have obtained for one and two quantum dot systems coupled to a plasmon, as well as an example involving fifty quantum dots. Section IV presents concluding remarks.

## II. THEORETICAL METHODS

## A. Lindblad master equation

The Lindblad master equation [8] for the time-dependent density matrix, $\rho(t)$, is

$$\frac{d}{dt}\rho(t) = \frac{1}{i\,h}\left[H(t),\rho(t)\right] + L\{\rho(t)\} \quad , \tag{1}$$

where $H$(t) is the system Hamiltonian (which may or may not include external driving in time) and $L\{\}$ is the Lindblad superoperator which acts on the density matrix as follows

$$L\{\rho\} = \sum_k C_k \rho C_k^+ - \frac{1}{2}\left(C_k^+ C_k \rho + \rho C_k^+ C_k\right) \quad , \tag{2}$$

where the sum is over the number of relevant dissipation or dephasing terms and $C_k$ are collapse operators to be specified below.

The Hamiltonians we study are those for one or more quantum dots coupled to a plasmonic mode [5-7],

$$H(t) = \hbar\omega_0 \sum_j \sigma_j^+ \sigma_j \; + \hbar\omega_{pl} b^+ b + \hbar \sum_j g_j (\sigma_j b^+ + \sigma_j^+ b) - \mu E(t) \qquad (3)$$

where $\sigma_j$ is a 2-state quantum dot lowering operator for quantum dot $j$, $b$ is a bosonic annihilation operator for the plasmonic mode and the $g_j$ are quantum dot/plasmon coupling rates. (Specific values for these and other parameters entering into the model will be given in the next section.) If an external field, $E$(t), is being considered, then the dipole operator is taken to be

$$\mu = \; d_0 \sum_j (\sigma_j^+ + \sigma_j) + d_{pl}(b^+ + b) \quad . \qquad (4)$$

(For simplicity we take the quantum dots to have the same excitation frequency and dipole moment parameters; in general, this is not necessary.)

The collapse operators $\{C_k\}$ correspond to spontaneous emission and pure dephasing for the quantum dots ($j = 1,2$), and plasmon damping (e.g., Ref. [11]) and are, respectively:

$$\sqrt{\gamma_1} \; \sigma_j, \; \sqrt{2\gamma_2} \sigma_j^+ \sigma_j \text{, and } \; \sqrt{\gamma_{pl}} \; b \; . \qquad (5)$$

With specification of a finite basis corresponding to two states per quantum dot and $N_{pl}$ plasmon states, the density matrix $\rho$ is either $2N_{pl}$ x $2N_{pl}$ for the case of one quantum dot or $4N_{pl}$ x $4N_{pl}$ for the case of two quantum dots and its elements can be written in operator form via (for the two quantum dot case)

$$\rho(t) = \sum_{s,q_1,q_2} \sum_{s',q_1',q_2'} \rho_{s,q_1,q_2,s',q_1',q_2'} \; (t) \; |s> |q_1> |q_2> < s'| < q_1'| < q_2'| \; , \quad (6)$$

where each $q_j = 0$ or 1 for ground or excited quantum dot states and $s = 0, 1, 2, \ldots, (N_{pl}\text{-}1)$.

We implement the density matrix approach described here using the convenient and freely available Quantum Toolbox in Python (QuTiP) [12, 13]. (For the low-intensity results here, $N_{pl} = 5$ suffices; for the high-intensity results, $N_{pl} = 15$.)

## B. Non-Hermitian model

Instead of solving the Lindblad master equation, Eq. (1), the non-Hermitian model involves solving a time-dependent Schrödinger equation (TDSE),

$$i\hbar \frac{d}{dt}\Psi(t) = H_c(t)\Psi(t) , \qquad (7)$$

where $H_c(t)$ is a complex Hamiltonian matrix or operator given by

$$H_c(t) = H(t) - \frac{i\hbar}{2}\sum_k C_k^+ C_k , \qquad (8)$$

where $H(t)$ is the system Hamiltonian, Eq. (2), and the $C_k$ are the collapse operators appearing in the Lindblad superoperator, Eq. (3), which take on the more explicit forms for our problems of interest in Eq. (5).

We have written the complex Hamiltonian in the form of Eq. (8) in order to draw its connection with the Lindblad master equation for the density matrix and in this form it can also be recognized as the effective Hamiltonian that enters into stochastic Schrödinger equation approaches as the first stage before any probabilistic collapses [14, 15]. However, one can easily re-express Eq. (8) as:

$$H_c(t) = \hbar\omega_0 \left(1 - i\frac{\Gamma}{2}\right)\sum_j \sigma_j^+ \sigma_j + \hbar\omega_{pl}(1 - i\frac{\gamma_{pl}}{2})\, b^+ b + \hbar g \sum_j (\sigma_j b^+ + \sigma_j^+ b) - \mu E(t), (9)$$

where $\Gamma = 2\gamma_2^* + \gamma_1$.

If the wave packet is written as

$$\Psi(t) = \sum_{s,q_1,q_2} \alpha_{s,q_1,q_2}(t)\, |s> |q_1> |q_2> \quad,\qquad\qquad (10)$$

then Eq. (6) yields first-order differential equations for the time derivatives $d\alpha_{s,q_1,q_2}(t)/dt$ and the contributions from the imaginary terms in Eqs. (8) or (9), i.e. the loss terms in these equations, are easily seen to be $-\left(\frac{\Gamma}{2}(q_1 + q_2) + s\gamma_{pl}\right)\alpha_{s,q_1,q_2}(t)$. There is therefore no loss for number states $q_1 = 0$ or $q_2 = 0$ or $s = 0$; also, the effective loss rate for each $s > 0$ plasmon state is $\frac{s\gamma_{pl}}{2}$, i.e. it increases with $s$. The non-Hermitian model results in a wave packet of sice $2N_{pl}$ for the case of one quantum dot. This is compared to the density matrix size of $2N_{pl}$ x $2N_{pl}$ for the full Lindblad master equation. Thus the non-Hermitian model results in significant computational savings while still including the effects of dissipation and dephasing.

## III. RESULTS

### A. Optical spectra

We first consider the case of one quantum dot interacting with a plasmonic system and employ the parameters of Shah et al. [5], which we list in Table I, along with other parameters used in our calculations. The parameters in this system are chosen to be consistent with the optical properties of two ellipsoidal gold nanoparticles of dimensions in the 20-30 nm range with a small

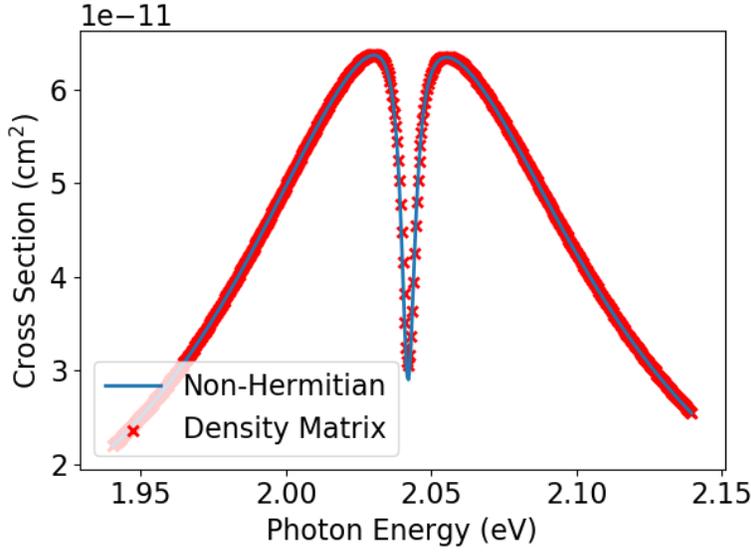

(4nm diameter) CdSe quantum dot placed within the 6 nm gap between the particles, all embedded within a medium of refractive index $n_{med} = 1.5$. Illumination is along the common, long axis and under such conditions it is possible to have a dipole or exciton-induced transparency effect [5, 16] wherein a sharp dip is superimposed on a broader, plasmonic lineshape.

Figure 1. Absorption spectra for one quantum dot interacting with a plasmon with the first parameter set of Table I.

We compute optical absorption spectra by considering the system in its ground state initially and exposing it to a short Gaussian pulse with sufficient energy content to describe the spectral region of interest,

$$E(t) = E_L \exp\left(-\left(\frac{t-t_c}{\tau_L}\right)^2\right)\cos\left(\omega_L t\right) \qquad . \qquad (11)$$

The optical spectrum is then computed from [5] (S.I. units)

$$\sigma(\omega) = \left(\frac{n_{med}\omega}{\epsilon_0 c}\right) Im[\alpha(\omega)] \qquad (12)$$

where the polarizability is given by

$$\alpha(\omega) = \frac{\int dt <\mu(t)> \exp\left(i\omega t\right)}{\sqrt{n_{med}} \int dt\, E(t)\exp\left(i\omega t\right)} \qquad , \qquad (13)$$

with the time integrals extending over times consistent with the system's response, the expectation value of the dipole moment (see Eq. (4)), $<\mu(t)>$, rising and falling to zero after the pulse.

Figure 1 displays the result of the non-Hermitian model (solid curve) and the full density matrix result based on solving the Lindblad density matrix (symbols). The agreement is quite good, with there being some small discrepancies in the narrowest region of the dipole-induced transparency. The dip is quite sensitive to quantum dot dephasing, as is evidenced in Figure 2 where we show how varying the dephasing rate from $\hbar\gamma_2^* = 0$ to $0.00508$ eV, keeping all other parameters as in Table I, can dramatically reduce the transparency. Figure 2(a) is the non-Hermitian model result and Fig. 2(b) is the density matrix result, again showing that the non-Hermitian model agrees well with the density matrix one.

Figure 3 displays optical spectra results for two quantum dots interacting with the plasmonic mode. Once again, very good agreement is seen between the non-Hermitian and density

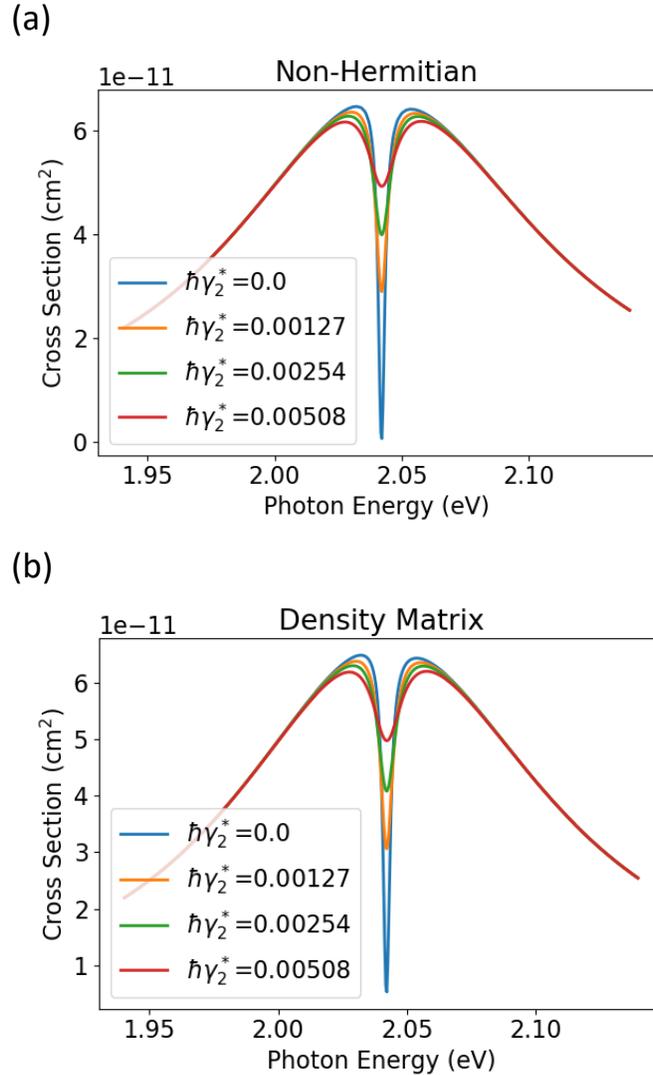

Figures 2(a) and 2(b): Absorption spectra for one quantum dot interacting with a plasmonic mode, now varying the quantum dot pure dephasing rate from $\hbar\gamma_2^* = 0$ to $0.00508$ eV. (a) The non-Hermitian model results, (b) the Lindblad density matrix results.

matrix approaches. In this case the dip is a little more pronounced than in the single quantum dot case of Figure 1. In fact, these results are consistent with a Tavis-Cummings picture [17] wherein the effect of n two-level systems interacting with a cavity (the plasmon) can be described by a single system interacting with the cavity with an effective system-cavity coupling factor of $\sqrt{n}\,g$. Indeed, if we carry out a calculation with one quantum dot interacting with the plasmon but employ dot-plasmon coupling $\sqrt{2}\,g$, the corresponding spectrum is virtually superimposable on the full two-dot results.

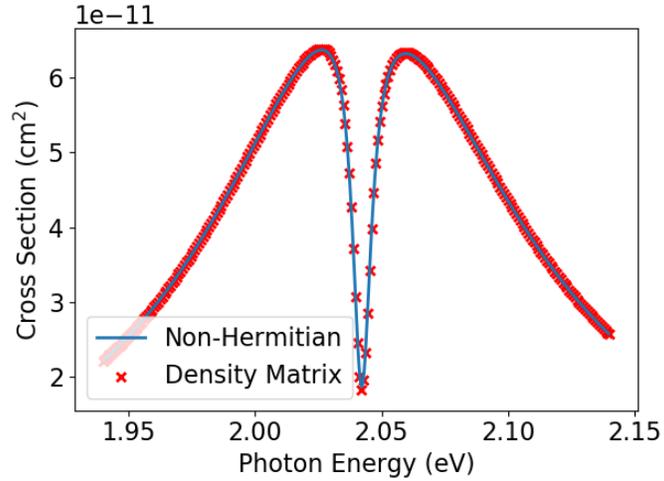

Figure 3. Absorption spectra for non-Hermitian (curve) and Lindblad density matrix (symbols) for the case of two quantum dots interacting with a plasmonic mode with the first set of parameters in Table I.

The optical spectra discussed above were inferred from Fourier transformation of the results from excitation with short pulses. The non-Hermitian model, however, can fail when the system is pumped continuous wave (CW) light, e.g., when the exponential term in Eq. (11) is set to unity. With both plasmon dissipation and dephasing operative, the system cannot achieve steady state populations owing to the loss terms leading to eventually complete loss of wave packet amplitude. Thus the model cannot describe the saturation effect of Ref. [5] that results when CW light is applied with ever increasing magnitudes of $E_L$. In this case, the Lindblad density matrix formalism leads to a diminishing of the transparency effect, i.e. the transparency dips become ever less deep to eventually being absent for sufficiently large magnitude $E_L$ [5]. Figure 4 illustrates the failure of the non-Hermitian model for CW excitation and $E_L$ = 1.4 x $10^{-6}$ atomic units (corresponding to an intensity on the order of 0.1 MW/cm$^2$, i.e. 100 times larger than that used in the calculations to generate Figs. 1-3). In this case we are continually applying the driving on the frequency or energy of the transparency dip, 2.042 eV. Whereas the low-intensity optical cross section from Figure 1 is ≈ 3 x $10^{-11}$ cm$^2$, the resulting cross section in this case is ≈ 6 x $10^{-11}$ cm$^2$, i.e. the dip has nearly been eliminated. In Fig. 4 we see that the density matrix calculations (symbols) yield a steady-state quantum dot number state expectation value, $< \sigma^+ \sigma >$, which is also the probability of excitation of the quantum dot, of slightly over 0.4, and the plasmon number state average, $< b^+ b >$, attains a value just under 0.1. In contrast, the non-Hermitian model (solid curves) is completely decayed away. The wave packet norm is shown as a green curve in Fig. 4 and the failure of the model can be correlated with it becoming significantly less than unity. The inset in Fig. 4 focuses on the shorter time behavior and shows that at least in this limit there is some agreement between the two approaches. We have experimented with a variety of "fixes", including renormalization after each time step and the introduction of gain terms. However, we have not yet found a suitable fix that would maintain the simplicity of a single wave packet propagation.

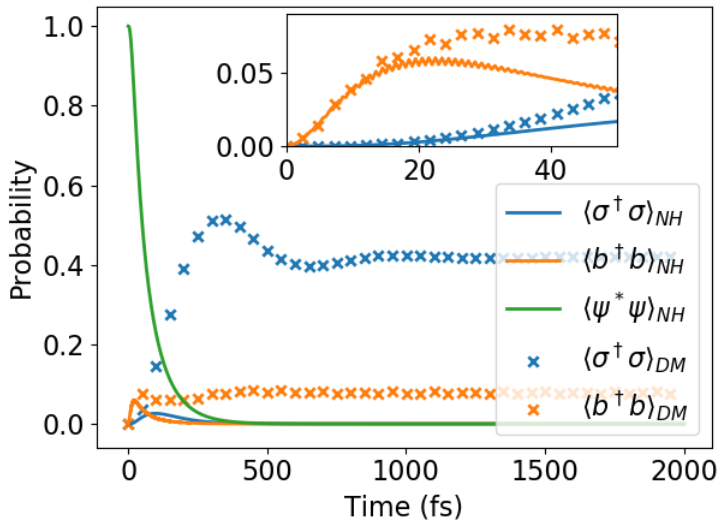

Figure 4. Illustration of the failure of the non-Hermitian model under constant, intense optical driving. The expectation values of the quantum dot and plasmon number states are shown as a function of time. The Lindblad master equation results (symbols) reach non-zero near steady state populations by 1200 fs, whereas the non-Hermitian model (curves) decays to zero. The very short time behavior displayed in the inset does show some agreement between the two approaches.

## B. Coherences and Entanglement

Rather than drive the system with a pulsed or CW laser, as in Sec. IIIA above, one can instead imagine the system to be already excited in some particular way and follow the subsequent time evolution in the absence of driving. For the case of two quantum dots coupled to a plasmon, we have found that if one dot is excited that energy transfer, mediated by the plasmon, can occur to the other quantum dot and that during the course of this process the two quantum dots can exhibit a reasonable degree of entanglement [6, 7, 18]. The non-Hermitian model can reproduce this type of behavior.

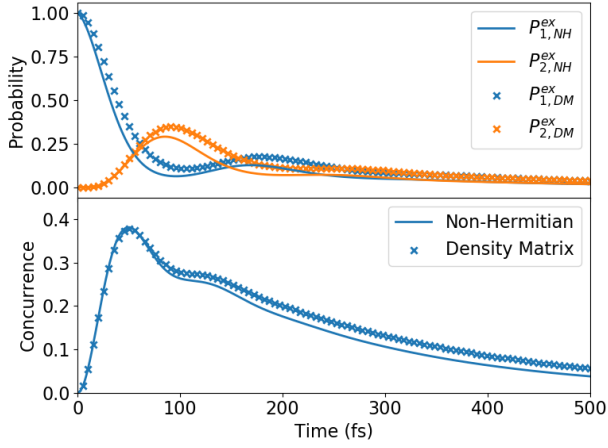

(a)

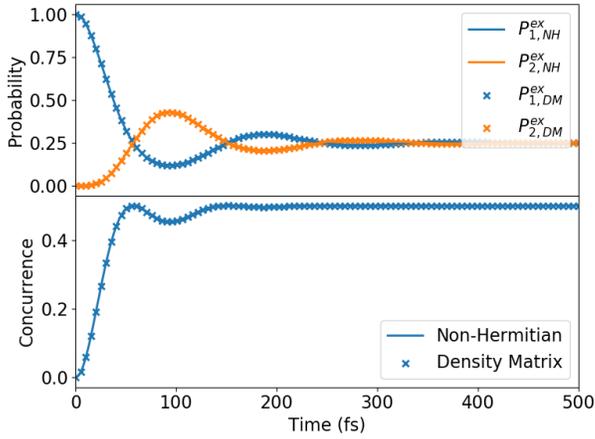

(b)

Figure 5. (a) Probabilities (upper panel) and concurrence (lower panel) exhibited by two quantum dots interacting with a plasmon (second parameter set of Table I; no external driving) when one quantum dot is initially excited and the rest of the system is cold. (b) Same as (a) except that the quantum dot pure dephasing is now set to zero.

behavior. This behavior does occur for the model parameters we have been using so far corresponding to those of Ref. [5]. However, the dynamics is somewhat uninteresting, exhibiting no transitory coherences. Somewhat more interesting behavior occurs when one uses the parameters of Ref. [18], which are designed to be consistent with a more complex plasmonic structure that exhibits gap plasmons. Employing these parameters (also listed in Table I), and now with no driving ($E_L = 0$) but setting the initial condition to be one of the quantum dots is in its excited state, we obtain the results in Figs. 5(a) and 5(b).

The upper panel of Fig. 5(a) shows that excited quantum dot probability comes down from unity as the second quantum dot excitation probability rises and there is also a subsequent, secondary coherence. However, while qualitatively correct, the non-Hermitian model (solid curves) does show discrepancies with the density matrix results (symbols). The lower panel of Fig. 5(b) shows the associated concurrence [19] of the two quantum dots, calculated by tracing out the plasmon quantum numbers to obtain a reduced two quantum dot density matrix and then performing the required computations [6, 7]. The concurrence is a measure of the degree of entanglement between the quantum dots, taking on a value of unity for maximal entanglement and 0 if the system is completely unentangled. Moderate values of entanglement, particularly at short times are achieved and the level of agreement between non-Hermitian and Lindblad density matrix concurrences is good, indicating (especially at the short times) that the non-Hermitian model can describe entanglement.

Interestingly, if we maintain the plasmon loss and spontaneous emission terms, but neglect the quantum dot dephasing, i.e., set $\gamma_2^* = 0$, we obtain the result in Fig. 5(b), with excellent agreement between the non-Hermitian and density matrix models, both in terms of probabilities and concurrences. Note that a steady state is established in the non-Hermitian model and it agrees with the density matrix one. In the Lindblad master equation approach, dephasing only affects the coherences while leaving the populations unchanged. Since the non-Hermitian Schrodinger equation approach deals directly with the probability amplitudes, it will affect both the coherences and populations simultaneously. It is only in the limit that dephasing goes zero that both approaches will agree quantitatively.

Dephasing is included in the non-Hermitian model, Eq. (9), as an imaginary part in some of the diagonal elements. This effectively acts as an additional source of dissipation, removing norm from system. In the full Lindblad master equation, however, dephasing does not just dissipate the number states; it also mixes the states and the effective dissipation due to dephasing is less than what the non-Hermitian model predicts. This is evidenced in Fig. 5(a) where the non-Hermitian dynamics are slightly below the density matrix dynamics.

We should point out that all the non-Hermitian model results of this subsection do not actually require numerical integrations since there is no driving and the dynamics is restricted to the one-excitation manifold. If the states in this one-excitation manifold are taken to be $|s = 1 >$ $|q_1 = 0 > |q_2 = 0 >$, $|s = 0 > |q_1 = 1 > |q_2 = 0 >$, $|s = 0 > |q_1 = 0 > |q_2 = (1 >$, the relevant 3 x 3 Hamiltonian matrix is

$$H_c = \hbar \begin{pmatrix} \omega_0 - i\,\gamma_s/2 & g & g \\ g & \omega_0 - i\,\Gamma/2 & 0 \\ g & 0 & \omega_0 - i\,\Gamma/2 \end{pmatrix} \quad , \qquad (14)$$

which is a generalization of the approach in the appendix of Ref. [7] (see also Ref. [20]) to include quantum dot dephasing and spontaneous emission. The eigenvalues and eigenvectors can readily be obtained and the dynamics of a particular initial condition within the one-excitation manifold can easily be obtained. It is also clear that this can easily be extended to studying the dynamics of many more quantum dots interacting with a plasmon mode. As an example, we consider fifty quantum dots in resonance with a single plasmonic mode with (i) homogeneous couplings (all dot/plasmon couplings equal to the value given in the lower parameter set of Table I (0.0167 eV)) and (ii) inhomogeneous couplings (dot/plasmon couplings randomly drawn from a normal distribution with mean and standard deviation 0.0167 eV). Figure 6 displays the average bipartite concurrence [19] for both cases with the initial condition being one quantum dot excited. The magnitudes of these concurrences are much smaller than the two quantum dot case. This is due to the phenomenon of monogomy of entanglement [21], where the amount of bipartite concurrence between any two systems of a set of systems decreases as the size of the set increases. Such states can still be maximally entangled (in a multi-partite sense), as has been shown for the W state [22]. In addition, Fig. 6 shows differences between the homogeneous and inhomogeneous coupling cases, with the transitory, maximum concurrences being somewhat larger for the homogeneous case and the inhomogeneous case exhibiting somewhat higher frequency oscillations. Due to the high efficiency of the non-Hermitian model in this single excitation manifold scenario, one could optimize the system parameters to construct other interesting multi-partite entangled states of potential relevance to error resiliency in quantum information applications, for example.

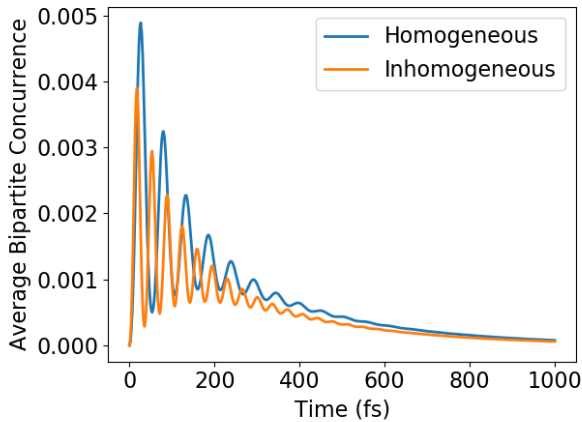

Figure 6. Average bipartite concurrence for fifty quantum dots interacting with a plasmonic system.

**IV. CONCLUDING REMARKS**

We have investigated a simple non-Hermitian model to describe quantum dot/plasmon interactions. We found that it yielded generally very good results in the linear optical excitation regime for models of one and two quantum dots coupled to a plasmonic structure. It led to poor results in the limit of CW driving as intensity was increased, however, and could not describe saturation effects in this limit. Nonetheless we were also able to show that the model could describe non-trivial coherences and entanglement in the un-driven case, i.e. scenarios wherein one imagines an optical process has already excited a quantum dot and energy transfer can occur via the plasmon to excite the other quantum dot. Such non-Hermitian models will be useful for studying cases involving many quantum dots coupled to plasmonic structures that cannot be easily simulated with more complete density matrix approaches.

In the future we plan to investigate approaches to remedying the failure of the non-Hermitian model in the high intensity CW case. One avenue is to exploit the fact that the model represents the first stage of the stochastic Schrödinger equation [12, 13]. Another avenue is to attempt to incorporate non-linearities explicitly into the equations that include aspects of gain, as in the optical Bloch equation work in Refs. [23, 24].

**ACKNOWLEDGMENTS**

This work was performed at the Center for Nanoscale Materials, a U.S. Department of Energy Office of Science User Facility, and supported by the U.S. Department of Energy, Office of Science, under Contract No. DE-AC02-06CH11357.

Table I.   Parameters used in the calculations unless otherwise specified in the text.

Optical spectra (Ref. [5]):

$\hbar\omega_0 = \hbar\omega_{pl} = \hbar\omega_L = 2.042$ eV
$\hbar g_1 = \hbar g_2 = 0.0108$ eV
$\hbar\gamma_1 = 268 \times 10^{-9}$ eV, $\hbar\gamma_2^* = 0.00127$ eV, $\hbar\gamma_{pl} = 0.150$ eV
$d_0 = 13.9$ Debye, $d_{pl} = 2990$ Debye
$E_L = 1.38 \times 10^{-7}$ atomic units (intensity 0.001 MW/cm$^2$)
$t_c = 50$ fs, $\tau_L = 10$ fs
Typical propagation time: 2500 fs

Coherences and entanglement (Ref. [18]):

$\hbar\omega_0 = \hbar\omega_{pl} = \hbar\omega_L = 1.44$ eV
$\hbar g_1 = \hbar g_2 = 0.0167$ eV
$\hbar\gamma_1 = 666 \times 10^{-9}$ eV, $\hbar\gamma_2^* = 0.0017$ eV, $\hbar\gamma_s = 0.033$ eV